\newif\ifels
\IfFileExists{elsarticle.cls}{\elstrue}{\elsfalse}
\ifels
  \documentclass[preprint,12pt]{elsarticle}
\else
  \documentclass[11pt,a4paper]{article}
  \usepackage[margin=2.4cm]{geometry}
  \usepackage[font=small,labelfont=bf]{caption}
\fi
\usepackage[utf8]{inputenc}
\usepackage{amsmath,amssymb}
\usepackage{graphicx}
\usepackage[colorlinks=true,linkcolor=blue,citecolor=blue,urlcolor=blue]{hyperref}

\newcommand{\Nch}{\langle dN_{ch}/d\eta\rangle}
\newcommand{\ssbar}{s\bar s}
\newcommand{\colfig}{0.6\textwidth}
\newcommand{\head}[1]{\section{#1}}

\newcommand{\papertitle}{Strangeness suppression, the $\phi$ meson and
$\Lambda\bar\Lambda$ spin correlations from exact strangeness conservation at the quark
level}
\newcommand{\paperabstract}{In small collision systems strange hadrons are suppressed,
and the suppression grows with the number of strange quarks. Exact strangeness
conservation in a finite volume accounts for this, but when it is imposed on hadrons it
leaves the $\phi$ meson unsuppressed, contrary to observation. We impose the
conservation law on the strange quarks in a causally connected cluster, before
hadronization, and count separately the combinations of a strange quark with the
antiquark created together with it. A pair created from the vacuum cannot form a
$\phi$, a pair created from a gluon can only if it loses its color. With one function of
multiplicity, the mean number of strange quarks per cluster, the model describes kaons,
$\phi$, $\Xi$ and $\Omega$ in pp and p--Pb collisions at the LHC; the data require that
about half of the same-pair combinations are excluded from the $\phi$, and reject both
an unsuppressed $\phi$ and a $\phi$ built only from different pairs at about five
standard deviations. The same-pair combinations are those that carry the spin correlation
observed in $\Lambda\bar\Lambda$ production, whose dilution is thereby tied to the
$\phi$ suppression and is predicted to decrease with multiplicity.}

\begin{document}

\ifels
\journal{Physics Letters B}
\begin{frontmatter}
\title{\papertitle}
\author[a,b]{P.~Castorina}
\author[c]{H.~Satz}
\address[a]{INFN, Sezione di Catania, Catania, Italy}
\address[b]{Faculty of Mathematics and Physics, Charles University, Prague, Czech Republic}
\address[c]{Fakult\"at f\"ur Physik, Universit\"at Bielefeld, Bielefeld, Germany}
\begin{abstract}\paperabstract\end{abstract}
\begin{keyword}
strangeness production \sep canonical suppression \sep $\phi$ meson \sep causality \sep spin correlations \sep small collision systems
\end{keyword}
\end{frontmatter}
\else
\title{\papertitle}
\author{P.~Castorina$^{1,2}$ and H.~Satz$^{3}$\\[2pt]
\small $^{1}$INFN, Sezione di Catania, Catania, Italy\\
\small $^{2}$Faculty of Mathematics and Physics, Charles University, Prague, Czech Republic\\
\small $^{3}$Fakult\"at f\"ur Physik, Universit\"at Bielefeld, Bielefeld, Germany}
\date{\today}
\maketitle
\begin{abstract}\noindent\paperabstract\end{abstract}
\fi

\head{Introduction}
In central nucleus-nucleus collisions the abundances of hadrons are those of an ideal
gas of resonances at a temperature $T\simeq155$ MeV~\cite{Andronic:2017pug}. In
elementary collisions strange hadrons are less abundant than this gas predicts. The
ALICE Collaboration has shown that in pp and p--Pb collisions their yields relative to
pions rise smoothly with the charged-particle multiplicity, towards the values of
central Pb--Pb collisions, and that the rise is stronger for hadrons with more strange
quarks~\cite{ALICE:2016fzo,ALICE:2015mpp,ALICE:2019avo}.

The suppression is described either by a factor $\gamma_s<1$ for each strange quark or
antiquark in the hadron~\cite{Letessier:1993hi,Becattini:2001fg}, or by the exact
conservation of strangeness in a finite correlation
volume~\cite{Rafelski:1980gk,Hagedorn:1984uy,Hamieh:2000tk,Kraus:2007hf,Kraus:2008fh}. The
second description has a physical basis. It was shown in Ref.~\cite{Castorina:2013mba}
that causality divides the hadronizing system into regions that cannot communicate, and
the correlation volume was identified with one such region. It has, however, a
well-known defect: the $\phi$ meson has no net strangeness and is not suppressed, while
the measured $\phi/\pi$ ratio rises with multiplicity~\cite{ALICE:2016sak}. The empirical
factor $\gamma_s^2$ for the $\phi$ works, but has no derivation. The defect is a
consequence of imposing the conservation law on hadrons, as done so far: at that level
the $\phi$ is a neutral particle, and its strange quark content plays no role.

In this letter we show that the defect is removed when the conservation law is imposed
on quarks rather than on hadrons, as suggested in Ref.~\cite{Castorina:2013mba}, provided
one takes into account how the strange quarks are produced. The same argument relates
the $\phi$ suppression to the spin correlation of $\Lambda\bar\Lambda$ pairs measured by
the STAR Collaboration~\cite{STAR:2025njp}.

\head{Causal clusters}
In a boost-invariant evolution with proper time $\tau$, partons come on shell at
$\tau_q$ and hadrons are formed at $\tau_h$. The largest region whose most distant
points can communicate between these two times has the proper longitudinal
extension~\cite{Castorina:2013mba}
\begin{equation}
d=\sqrt{\frac{\tau_h}{\tau_q}}\,(\tau_h-\tau_q). \label{eq:d}
\end{equation}
Strange quarks are created in pairs, by the breaking of a color string or by the
splitting of a gluon. Both processes are local, and the two members of a pair cannot
separate by more than the causal range before hadronization. The partner of each
strange quark is therefore found within the distance $d$. We call this region a
cluster and take it to be a sphere of radius $R=d/2$ and volume $V$. In every cluster
the numbers of strange quarks and antiquarks are equal.

\head{Pair distribution}
Without this constraint the numbers of $s$ and $\bar s$ in a cluster are independent
Poisson variables with mean $x/2$, so that $x$ is the mean number of strange quarks
plus antiquarks. With the constraint, the number $n$ of pairs is distributed as (see
Supplemental Material)
\begin{equation}
P(n)=\frac{1}{I_0(x)}\,\frac{(x/2)^{2n}}{(n!)^2}, \label{eq:P}
\end{equation}
where $I_k$ is the modified Bessel function of order $k$; this is the zero-difference
term of the Skellam distribution~\cite{Skellam:1946}. Its factorial moments follow from
the series of $I_k$,
\begin{equation}
\big\langle n(n-1)\cdots(n-k+1)\big\rangle=\Big(\frac{x}{2}\Big)^{k}\,S_k,\qquad
S_k\equiv\frac{I_k(x)}{I_0(x)} . \label{eq:fact}
\end{equation}
A hadron with $k$ strange quarks, and no strange antiquark, can be formed in a number
of ways proportional to this moment. Without the constraint the number is $(x/2)^k$.
The hadron is therefore suppressed, relative to the limit of a large cluster, by
$S_k$: $S_1$ for $K$ and $\Lambda$, $S_2$ for $\Xi$, $S_3$ for $\Omega$. The
proportionality constant, which contains the probability that the quarks form the
hadron, is assumed not to depend on the size of the cluster, and is fixed by the limit of
a large cluster.

\head{Why the $\phi$ is suppressed}
For a hadron made of a strange quark and a strange antiquark the relevant quantity is
$\langle n^2\rangle$, and Eq.~(\ref{eq:P}) gives
\begin{equation}
\langle n^2\rangle=\Big(\frac{x}{2}\Big)^2 \label{eq:n2}
\end{equation}
exactly, as without the constraint. Exact conservation alone does not suppress hidden
strangeness, at the quark level as at the hadron level: the partner of each strange
quark is always present.

Equation~(\ref{eq:P}) fixes the number of pairs produced in a cluster. At hadronization,
a strange quark can combine with any of the $n$ antiquarks: with its own partner, or
with the partner of another quark. The two cases differ in the quantum numbers of the
combination, which are those of the production process for the first and are
uncorrelated for the second. Accordingly, Eq.~(\ref{eq:n2}) is the sum of two terms (see
Supplemental Material),
\begin{equation}
\langle n^2\rangle=\underbrace{\langle n(n-1)\rangle}_{\text{different pairs}}
+\underbrace{\langle n\rangle}_{\text{same pair}} . \label{eq:split}
\end{equation}
Divided by $(x/2)^2$, the two terms are $S_2$ and $(2/x)\,S_1=1-S_2$, by the recurrence
$I_0(x)-I_2(x)=(2/x)\,I_1(x)$. The second term counts the combinations of a strange
quark with the antiquark created together with it. Whether such a pair can become a $\phi$ depends on how it was
produced. Created from the vacuum, it has $J^{PC}=0^{++}$, with parallel spins and one
unit of orbital angular momentum~\cite{LeYaouanc:1972vsx}, while the $\phi$
has $J^{PC}=1^{--}$ and none: it cannot be a $\phi$. Created from a gluon, it has the
quantum numbers of the $\phi$ but is a color octet. It can form a $\phi$ only after its
color has been neutralized by the exchange of soft gluons with the surrounding partons,
as in the color-octet mechanism of quarkonium production. This step
is nonperturbative; moreover, for strange quarks, whose mass is of the order of the QCD
scale, the relative weight of the two production mechanisms cannot be computed
perturbatively either. We therefore introduce a parameter and write
\begin{equation}
C_\phi=S_2+w\,(1-S_2), \label{eq:Cphi}
\end{equation}
where the first term comes from different pairs, Eq.~(\ref{eq:fact}), and $w$ is the
fraction of same-pair combinations that can form a $\phi$. For $w=0$ the $\phi$ behaves
as a hadron with two strange quarks; for $w=1$ it is not suppressed, as in the canonical
description of hadrons. The suppression of the $\phi$ is thus not a consequence of the
conservation law, but of the quantum numbers of the pairs.

\begin{figure}[t]
\centering
\includegraphics[width=\colfig]{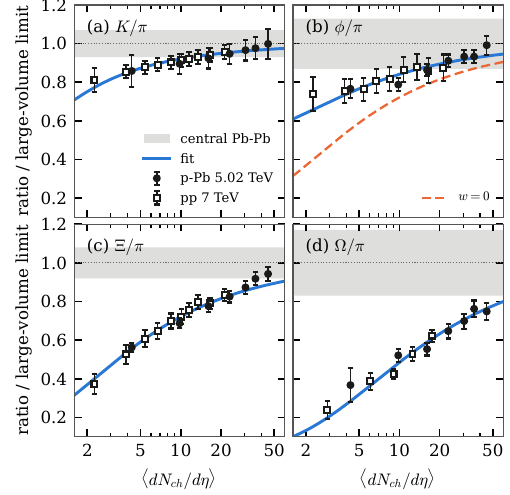}
\caption{Ratios to pions in p--Pb collisions at $\sqrt{s_{NN}}=5.02$
TeV~\cite{ALICE:2016sak,ALICE:2015mpp} and in pp collisions at $\sqrt s=7$
TeV~\cite{ALICE:2018pal}, divided by the large-volume limits and by the fitted
normalization factors, versus charged-particle multiplicity. Solid lines: $S_1$, $C_\phi$, $S_2$ and
$S_3$ with Eq.~(\ref{eq:fit}). Dashed line in (b): $\phi$ built from different pairs
only, $w=0$. Shaded band: the ratio measured in central Pb--Pb collisions, to which the
large-volume limit is fixed, with its uncertainty. The normalization factors are 0.96, 0.99, 1.23 and 0.96 for $K$, $\phi$,
$\Xi$ and $\Omega$ in p--Pb collisions, and 0.96, 1.03, 1.15 and 0.90 in pp
collisions.}
\label{fig:ratios}
\end{figure}

\head{Comparison with data}
We use the ALICE measurements of $K/\pi$, $\phi/\pi$, $\Xi/\pi$ and $\Omega/\pi$ in
p--Pb collisions at $\sqrt{s_{NN}}=5.02$ TeV, in seven multiplicity classes with
$N\equiv\Nch$ from 4.3 to 45~\cite{ALICE:2016sak,ALICE:2015mpp}, and in pp collisions at
$\sqrt s=7$ TeV, in ten classes with $N$ from 2.3 to 21~\cite{ALICE:2018pal}: 62 points
in all. The model gives the suppression of each ratio with respect to its value in a
large cluster, not the value itself: the absolute value depends on the probability that
the quarks form a given hadron, which is not calculated here and, being independent of
the size of the cluster, cancels in the ratio to the large-cluster limit. We take the
latter from experiment: the limits are
fixed to the ratios measured in central Pb--Pb collisions at
$\sqrt{s_{NN}}=2.76$ TeV~\cite{ALICE:2013mez,ALICE:2013xmt,ALICE:2014jbq}, $K/\pi=0.149$,
$\phi/\pi=9.4\times10^{-3}$, $\Xi/\pi=4.99\times10^{-3}$ and
$\Omega/\pi=8.9\times10^{-4}$, and are not fitted. Systematic
uncertainties that are common to all multiplicity classes are described by one
normalization factor for each ratio and each collision system, constrained to unity
within the correlated uncertainty (6 to 12\%). In Fig.~\ref{fig:ratios} the data are divided by the fitted
factors, which are given in the caption. The remaining, point-to-point
uncertainties are the statistical ones and the part of the systematic uncertainty that
is uncorrelated between classes. The latter is given in the experimental papers for the
p--Pb data~\cite{ALICE:2016sak,ALICE:2015mpp}; for the pp data it is obtained from the
published breakdown of the systematic uncertainties~\cite{ALICE:2018pal}, as described
in Sec.~S4 of the Supplemental Material. We parametrize $x=a\,N^{b}$, with
the same function for both systems (for the corresponding cluster size see
Sec.~\ref{sec:size}). The procedure is described in the Supplemental
Material.

The fit gives
\begin{equation}
a=1.54\pm0.13,\qquad b=0.63\pm0.06,\qquad w=0.43\pm0.09, \label{eq:fit}
\end{equation}
with $\chi^2=37.9$ for 59 degrees of freedom; $a$, $b$ and $w$ are the only parameters
of the model. They determine the dependence on multiplicity of the eight sets of data;
the absolute level of each set is tested within its normalization uncertainty. The
value of $\chi^2$ is well below the number of degrees of freedom, which indicates that
the point-to-point uncertainties are, if anything, overestimated. The quality of the fit
does not depend on this: with these uncertainties reduced by 30\% the fit gives
$\chi^2=60.7$ for the same 59 degrees of freedom and the same values of $a$, $b$ and
$w$. A cluster contains on average 2.6 strange
quarks and antiquarks at $N=2.3$ and 17 at $N=45$. Figure~\ref{fig:ratios} shows the
result. The two collision systems follow the same curves: if $a$ and $b$ are fitted
separately for pp and p--Pb collisions, $\chi^2$ decreases by 0.1 for two more
parameters. Between $N=2.3$ and $N=45$ the factors $S_1$, $S_2$ and $S_3$ go from 0.76,
0.40 and 0.15 to 0.97, 0.89 and 0.76, and $C_\phi$ from 0.66 to 0.94. The normalization
factors are within their uncertainties except for the $\Xi$, whose yield at high
multiplicity exceeds the Pb--Pb reference.

\begin{figure}[t]
\centering
\includegraphics[width=\colfig]{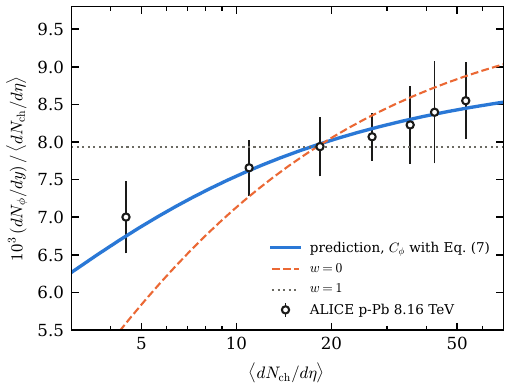}
\caption{$\phi$ yield per charged particle in p--Pb collisions at
$\sqrt{s_{NN}}=8.16$ TeV~\cite{ALICE:2021uyz} versus charged-particle multiplicity.
Solid line: prediction with the parameters of Eq.~(\ref{eq:fit}), fixed by the data of
Fig.~\ref{fig:ratios}, with free normalization. Dashed and dotted lines: $w=0$ and
$w=1$, each with its own normalization.}
\label{fig:816}
\end{figure}

The value of $w$ is determined by the $\phi$. An unsuppressed $\phi$, $w=1$, gives
$\chi^2=70.4$, and a $\phi$ built from different pairs only, $w=0$, gives $\chi^2=60.7$:
they are rejected at 5.7 and 4.8 standard deviations. The $\phi$ is suppressed, but about half
as much as a hadron with two strange quarks [Fig.~\ref{fig:ratios}(b)]. In the picture of
the previous section, $w$ is the product of the fraction of pairs created by gluon
splitting and of the probability that such a pair becomes a color singlet. 
A test of this result is provided by the $\phi$ yield measured in p--Pb collisions at
$\sqrt{s_{NN}}=8.16$ TeV, in seven multiplicity classes with $N$ from 4.5 to
53~\cite{ALICE:2021uyz}. The pion yields are not available at this energy, and these
data are not used in the fit. With $a$, $b$ and $w$ fixed by Eq.~(\ref{eq:fit}), the
model predicts the dependence on multiplicity of the $\phi$ yield per charged
particle, if the number of pions per charged particle is constant; only the overall
normalization is free. The comparison is shown in Fig.~\ref{fig:816}. The prediction
gives $\chi^2=0.5$ for 6 degrees of freedom, against 6.7 for an unsuppressed $\phi$ and
11.5 for a $\phi$ built from different pairs only. The Supplemental Material gives the
same comparison with the pions per charged particle taken from the data at 5.02 TeV: the
conclusion is the same.

The $\Lambda$ has the same
factor as the kaon. In pp collisions at $\sqrt s=7$ TeV the ratio $\Lambda/K$ is
constant within 2\% for $N>5$ and decreases by 20\% at the lowest
multiplicity~\cite{ALICE:2018pal}; it is constant within uncertainties at 13
TeV~\cite{ALICE:2019avo}. The decrease at low multiplicity, and the behavior of baryons
and light nuclei, involve the corresponding constraint for the light flavors and are
discussed in a separate paper.

\head{Cluster size}\label{sec:size}
For strange quarks of mass $m_s$ and degeneracy $g=6$,
\begin{equation}
x=2Vn_s,\qquad n_s=\frac{g}{2\pi^2}\,m_s^2\,T\,K_2\!\Big(\frac{m_s}{T}\Big),
\end{equation}
with $K_2$ the modified Bessel function of the second kind. For $T=155$ MeV and
$\tau_q=1$ fm, Eqs.~(\ref{eq:d}) and~(\ref{eq:fit}) give, between $N=2.3$ and $N=45$,
a radius from 1.0 to 2.0 fm and $\tau_h$ from 2.4 to 3.2 fm for a current mass
$m_s=100$ MeV, and a radius from 1.5 to 2.9 fm and $\tau_h$ from 2.8 to 3.9 fm for a
constituent mass $m_s=460$ MeV. In both cases the hadronization time is close to the
2--3 fm estimated in Ref.~\cite{Castorina:2013mba} from data at lower energies and
grows slowly with multiplicity.

\head{Spin correlation of $\Lambda\bar\Lambda$ pairs}
The STAR Collaboration has measured a relative polarization
$0.181\pm0.035\pm0.022$ for $\Lambda\bar\Lambda$ pairs at small separation in rapidity and azimuthal angle in pp
collisions at $\sqrt s=200$ GeV, and no signal for widely separated pairs or for
$\Lambda\Lambda$ pairs~\cite{STAR:2025njp}. The result is interpreted as the survival of
the parallel spins of an $\ssbar$ pair created from the vacuum.

In a cluster, a $\Lambda$ and a $\bar\Lambda$ carry this correlation only if their
strange quark and antiquark belong to the same pair; a pair from a gluon has parallel
spins as well. The fraction of such combinations is the ratio of the second term of
Eq.~(\ref{eq:split}) to the total,
\begin{equation}
D=\frac{\langle n\rangle}{\langle n^2\rangle}=\frac{2}{x}\,S_1=1-S_2 , \label{eq:D}
\end{equation}
where the last step uses the recurrence $I_0(x)-I_2(x)=(2/x)\,I_1(x)$. With
Eq.~(\ref{eq:Cphi}),
\begin{equation}
1-C_\phi=(1-w)\,D . \label{eq:rel}
\end{equation}
The same-pair combinations give the spin correlation; a fraction $1-w$ of them is lost
to the $\phi$. The suppression of the $\phi$ and the dilution of the spin correlation
are therefore proportional at all multiplicities.

\begin{figure}[t]
\centering
\includegraphics[width=\colfig]{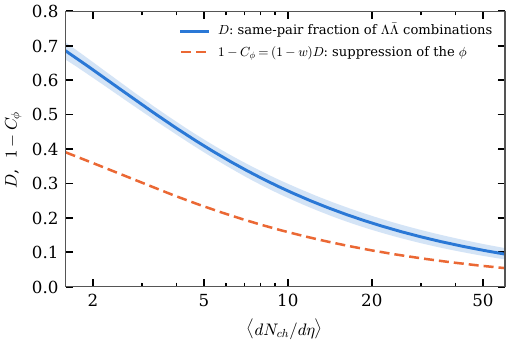}
\caption{Solid line: fraction $D$ of $\Lambda\bar\Lambda$ combinations whose strange
quark and antiquark come from a single pair, Eq.~(\ref{eq:D}), from the fit of
Fig.~\ref{fig:ratios}, with its $1\sigma$ uncertainty. Dashed line: suppression of the
$\phi$, $1-C_\phi$, Eq.~(\ref{eq:rel}).}
\label{fig:D}
\end{figure}

Figure~\ref{fig:D} shows $D$ obtained from the fit to the strange hadron yields. It
falls from $0.44\pm0.02$ at $N=4.3$ to $0.21\pm0.02$ at $N=16$ and $0.11\pm0.02$ at
$N=45$. The prediction is that the $\Lambda\bar\Lambda$ spin correlation decreases with
multiplicity in proportion to $D$, and vanishes in central nucleus-nucleus collisions.
The polarization transfer from the strange quark to the $\Lambda$ and the feed-down
enter as a common factor, which largely cancels in the ratio of the correlation at two
multiplicities; the multiplicity dependence is therefore a more robust test than the
absolute value.

The absolute value is less certain. For pp collisions at $\sqrt s=200$ GeV,
$N\simeq3$~\cite{STAR:2008med} and, if $x$ depends only on multiplicity, $D=0.53$. If
all the pairs are in the spin-triplet state and transfer their spin entirely to the
$\Lambda$ and the $\bar\Lambda$, the relative polarization is $1/3$; with the SU(6) wave
function and the feed-down fractions of PYTHIA, Ref.~\cite{STAR:2025njp} estimates a
maximum of 0.096. The dilution then gives 0.05, about three standard deviations below
the measured value. This estimate of the maximum by the STAR Collaboration is, however, itself two
standard deviations below the measured value: even without any dilution ($D=1$) it
would not reproduce the data.

Notice that with the bare factor $1/3$, i.e.\ without the reduction due to feed-down, the
dilution gives 0.18, close to the measured value. The
selection of nearby pairs, which favors same-pair combinations, would increase both
estimates. The absolute comparison thus depends on the polarization transfer more than
on $D$.

\head{Conclusions}
Exact strangeness conservation in a causal cluster, imposed on the quarks before
hadronization, reproduces the multiplicity dependence of kaons, $\Xi$ and $\Omega$ in
pp and p--Pb collisions with one function. For the $\phi$ the conservation law is not
enough; what matters is whether a strange quark and the antiquark created with it can
form a vector, color-singlet state. The data show that about half of these
combinations cannot, which is what one expects if they carry the quantum numbers of the
vacuum. The same combinations preserve the spin correlation of a single pair in
$\Lambda\bar\Lambda$ production. The suppression of the $\phi$ and the dilution of the
spin correlation are tied by Eq.~(\ref{eq:rel}), which can be tested by measuring the
$\Lambda\bar\Lambda$ spin correlation as a function of multiplicity.

\end{document}